\documentclass[sigconf]{acmart}

\AtBeginDocument{%
  }

\setcopyright{acmcopyright}
\copyrightyear{2018}
\acmYear{2018}
\acmDOI{XXXXXXX.XXXXXXX}

\acmConference[Conference acronym 'XX]{Make sure to enter the correct
  conference title from your rights confirmation emai}{June 03--05,
  2018}{Woodstock, NY}
\acmPrice{15.00}
\acmISBN{978-1-4503-XXXX-X/18/06}

\usepackage{subfigure}
\usepackage{makecell}

\usepackage{hyperref}

\usepackage[english]{babel}
\addto\extrasenglish{%
}

\usepackage{hyphenat}

\usepackage{forloop}

\newsavebox\MyBreakChar%
\sbox\MyBreakChar{}
\newsavebox\MySpaceBreakChar%
\sbox\MySpaceBreakChar{\hyp}
\makeatletter%
\newcommand*{\BreakableChar}[1][\MyBreakChar]{%
  \leavevmode%
  \discretionary{\usebox#1}{}{}%
}%
\makeatother

\newcounter{index}%
\newcommand{\AddBreakableChars}[1]{%
  \StrLen{#1 }[\stringLength]%
  \forloop[1]{index}{1}{\value{index}<\stringLength}{%
    \StrChar{#1}{\value{index}}[\currentLetter]%
    \IfStrEqCase{\currentLetter}{%
        {*}{\currentLetter\BreakableChar[\MyBreakChar]}%
        {/}{\currentLetter\BreakableChar[\MyBreakChar]}%
        {+}{\currentLetter\BreakableChar[\MyBreakChar]}%
        {\&}{\currentLetter\BreakableChar[\MyBreakChar]}%
    }[\currentLetter]%
  }%
}%

\def\fVR{\includegraphics[height=1em]{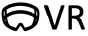}}
\def\fDesktop{\includegraphics[height=1em]{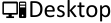}}
\def\fLinear{\includegraphics[height=1em]{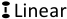}}
\def\fBranch{\includegraphics[height=1em]{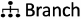}}

\def\fVRBold{\includegraphics[height=1em]{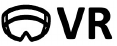}}
\def\fDesktopBold{\includegraphics[height=1em]{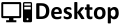}}
\def\fLinearBold{\includegraphics[height=1em]{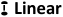}}
\def\fBranchBold{\includegraphics[height=1em]{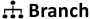}}

\def\fvrLinearBold{\fVRBold{}\textbf{ +}\fLinearBold{}}

\def\fvrBranchBold{\fVRBold{}\textbf{ +}\fBranchBold{}}

\def\fdesktopLinearBold{\fDesktopBold{}\textbf{ +}\fLinearBold{}}

\def\fdesktopBranchBold{\fDesktopBold{}\textbf{ +}\fBranchBold{}}

\def\one{\includegraphics[height=1em]{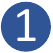}}
\def\two{\includegraphics[height=1em]{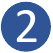}}
\def\three{\includegraphics[height=1em]{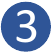}}
\def\four{\includegraphics[height=1em]{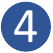}}

\def\desktopLinearT{\textsc{Desktop+\BreakableChar{}Linear}}
\def\desktopBranchT{\textsc{Desktop+\BreakableChar{}Branch}}
\def\vrLinearT{\textsc{VR+\BreakableChar{}Linear}}
\def\vrBranchT{\textsc{VR+\BreakableChar{}Branch}}

\def\VRT{\textsc{VR}}
\def\DesktopT{\textsc{Desktop}}

\def\LinearT{\textsc{Linear}}
\def\BranchT{\textsc{Branch}}

\begin{document}

\title[]{
Evaluating Navigation and Comparison Performance of Computational Notebooks on Desktop and in Virtual Reality}




\author{Sungwon In}
\affiliation{%
  \institution{Virginia Tech}
  \city{Blacksburg}
  \country{United States}}
\email{sungwoni@vt.edu}

\author{Eric Krokos}
\affiliation{%
  \institution{Department of Defense}
  \country{United States}}

\author{Kirsten Whitley}\orcid{0000-0003-1356-326X}
\affiliation{%
  \institution{Department of Defense}
  \country{United States}}

\author{Chris North}
\affiliation{%
  \institution{Virginia Tech}
  \city{Blacksburg}
  \country{United States}}
\email{north@cs.vt.edu}

\author{Yalong Yang}
\affiliation{%
  \institution{Georgia Tech}
  \city{Atlanta}
  \country{United States}}
\email{yalong.yang@gatech.edu}

\renewcommand{\shortauthors}{Trovato et al.}

\begin{abstract}
The computational notebook serves as a versatile tool for data analysis. However, its conventional user interface falls short of keeping pace with the ever-growing data-related tasks, signaling the need for novel approaches. 
With the rapid development of interaction techniques and computing environments, there is a growing interest in integrating emerging technologies in data-driven workflows. Virtual reality, in particular, has demonstrated its potential in interactive data visualizations. 
In this work, we aimed to experiment with adapting computational notebooks into VR and verify the potential benefits VR can bring. We focus on the navigation and comparison aspects as they are primitive components in analysts' workflow. 
To further improve comparison, we have designed and implemented a Branching\&Merging functionality. We tested computational notebooks on the desktop and in VR, both with and without the added Branching\&Merging capability. We found VR significantly facilitated navigation compared to desktop, and the ability to create branches enhanced comparison.
\end{abstract}



\begin{teaserfigure}
  \includegraphics[width=\textwidth]{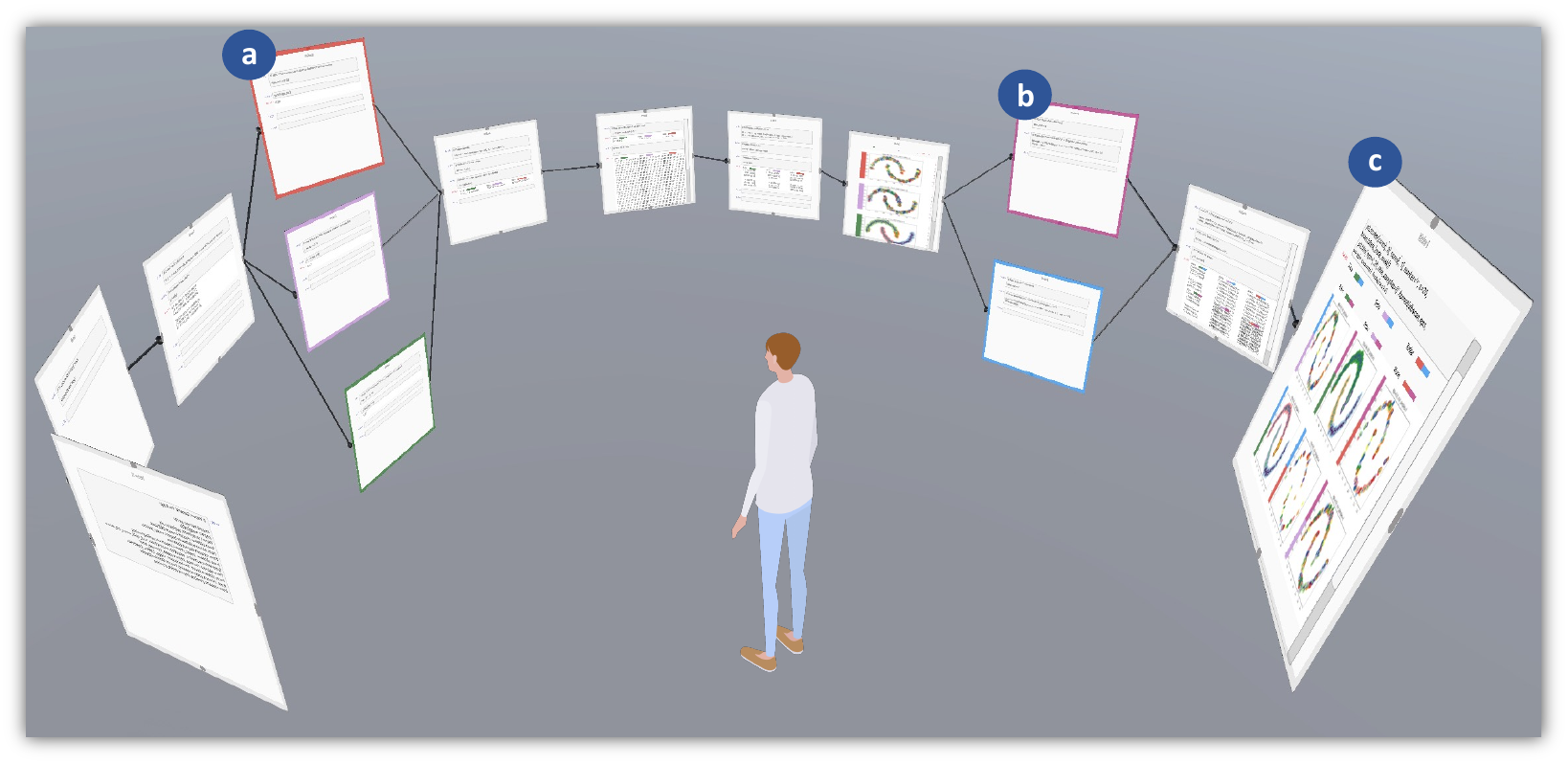}
  \caption{A demonstration of the immersive computational notebook system featuring the branching capabilities. The user can work with windows containing multiple cells, which are interconnected to indicate execution order. 
  (a) and (b) shows branches created at different locations of the notebook for testing multi-level hypotheses.
  (c) subsequently merge these branches for streamlined result comparisons, with six results---three branches at (a) $\times$ two branches at (b)---presented in the output of the window.}
  \Description{A showcase of the immersive computational notebook system highlights its branching functionalities. Users have the ability to interact with windows comprising multiple cells, which are interlinked to signify the sequence of execution. (a) and (b) illustrate branches that have been initiated at various points within the notebook, aimed at evaluating multi-level hypotheses. Multiple colors were used to indicate each window in the branch containing a different kernel. (c) Following the creation of these branches, they can be merged to facilitate streamlined comparisons of results.}
  \label{fig:teaser}
  \vspace{1.0em}
\end{teaserfigure}

\maketitle

\section{Introduction}
\label{sec:intro}

The computational notebook has gained substantial popularity across diverse domains due to its versatility, characterized by its seamless integration of code, documentation, and visual outputs in a unified interface, ease of sharing and replication, and interactive features with real-time feedback.
Data analysts leverage these capabilities for tasks ranging from constructing analytical pipelines to debugging and comparative analysis~\cite{chattopadhyay2020s}.

However, as data analysis grows in complexity, the standard computational notebook's user interface shows limitations in supporting the aforementioned tasks.
Specifically, \textit{navigating} through extensive notebooks becomes increasingly challenging, complicating tasks like identifying issues in the analytical pipeline or refactoring code{~\cite{kery2018story}}. 
Analysts often have to scroll up and down through lengthy sections multiple times, relying heavily on their working memory for off-screen content, leading to high context-switching costs{~\cite{kery2017variolite}}.
Additionally, modern data analysis often involves \textit{comparisons}, such as determining optimal parameters for analytical models. Common practices, like duplicating code or notebook, complicate code and execution management by introducing non-linearities{~\cite{deline2012debugger}}. 
While literature identifies other challenges with computational notebooks{~\cite{chattopadhyay2020s}}, this project focuses on enhancing \textit{navigation} and \textit{comparison}, which are primitive components of data analysts' workflows{~\cite{raghunandan2023code}}.

Efforts to improve navigation and comparison in computational notebooks within the desktop environment exist{~\cite{weinman2021fork}}. 
However, the inherent limitations of desktop display and interaction paradigms restrict the spatial presentation and interaction modes with computational notebooks. 
Thus, we explore opportunities offered by emerging technologies, specifically virtual reality (VR). 
VR headsets allow interaction with 2D and 3D graphics and interfaces in an expansive space, introducing new human-computer interaction possibilities for non-linear notebooks. 
Preliminary research in using VR for data science shows promising benefits.
For example, VR allows analysts to use its large space as an external memory layer with spatial semantic meanings to better support information retrieval{~\cite{lisle_sensemaking_2021,davidson2022exploring,reiske2023multi}}. 
Physical interactions in VR, including natural walking and embodied gestures, provide rapid information access and command execution{~\cite{in2023table,yang2020embodied,tong2023towards,huang2023embodied}}.
Consequently, our overarching research question is: \textbf{Can VR's spatial and embodied nature enhance navigation and comparison in computational notebooks?}

To approach the proposed research question, we first adapted the computational notebook for the VR environment. 
Specifically, we introduced an additional hierarchy layer to facilitate notebook content management, adopted a curved layout for content placement, and designed gesture-based interactions, including a branch\&merge gesture to assist comparison.
To systematically verify and understand the potential benefits of virtual reality, we conducted a controlled user study to compare computational notebooks on the desktop and in VR, both with and without branch\&merge capability.
The study task comprised two phases: first, participants were asked to \textit{navigate} through a presented computational notebook, identifying and rectifying deliberate issues; second, they were required to determine optimal parameter values through \textit{comparisons}. 
Pre-configured codes were provided to reduce the need for constructing a notebook from scratch, enabling participants to focus on \textit{navigation} and \textit{comparison} tasks.
We noticed that participants encountered significant challenges in editing text while in VR.
After excluding the text editing time from all conditions, we found that VR had better navigation performance than the desktop, and the branch\&merge functionality significantly facilitated the comparison process. 
The contributions of this work are twofold: 1) the adaptation of computational notebooks from desktop to VR, and 2) empirical knowledge about using computational notebooks in VR.

\vspace{-1em}
\section{Related Work}
\label{sec:related_work}

\subsection{Challenges in Computational Notebooks}
\label{sec:related_work_challenges}

Drawing from Knuth's literate programming paradigm~\cite{knuth1984literate}, computational notebooks seek to construct a computational narrative, enhancing analysts' efficiency in iterative data science tasks by amalgamating visuals, text, and analytical insights into an integrated document ~\cite{rule2018exploration}.
Numerous implementations, such as Jupyter Notebook~\cite{JupyterNotebook}, DataBricks~\cite{DataBricks}, Apache Zeppelin~\cite{Zeppelin}, and CarbidAlpha~\cite{CarbideAlpha}, have been developed.
Yet, as data analyses evolve in complexity, these platforms present challenges in supporting the increasingly intricate data science workflows.
In 2015 and 2020, Jupyter, a leading computational notebook application, conducted user surveys to illuminate these issues{~\cite{JupyterSurvey}}, receiving feedback mainly on system functionalities like encompassing performance, sharing capabilities, version control, and enriched documentation.
Certain user experience concerns also emerged, like content collapse, progress indicators, and global search.
Aligning with this endeavor, Chattopadhyay et al.~\cite{chattopadhyay2020s} undertook a rigorous exploration involving 156 data science professionals to systematically unravel the pain points, needs, and opportunities with computational notebooks.
Their study spotlighted nine pivotal challenges that not only add operational hurdles but also layer on complexities detrimental to analytical workflows.

Our study particularly addresses challenges tied to the visual and interactive facets of computational notebooks, with an emphasis on exploration and code management.
As underscored by previous research~\cite{head2019managing,rule2018exploration}, during exploratory phases, analysts often prioritize flexibility and speed over clarity and sustainability, leading to long and ``messy'' notebooks.
A prevalent practice includes cloning variables, code segments, or entire notebooks as an informal versioning method, bypassing standard tools~\cite{kery2018story,tabard2008individual}.
As a result, unintentional modifications and deletions in notebooks make data analysis error-prone and laborious{~\cite{rule2018exploration,kery2018story}}, and locating specific elements becomes more challenging with increasing complexity, as echoed in the Jupyter survey{~\cite{rule2018exploration,tabard2008individual,kery2018story}}.
Taking cues from prior research, we focus on supporting \textit{navigation} and \textit{comparison} of computation notebooks by leveraging the display and interaction capabilities of immersive environments.

\subsection{Navigation and Comparison in Computational Notebooks}
Improving the navigation experience has historically been a central focus in user interface and interaction design, resulting in various techniques for diverse applications.
Among them, notably, techniques like focus+context and overview+detail have gained significant traction.
Cockburn et al.~\cite{cockburn2009review} provided an exhaustive review of these techniques.
Focus+context techniques like the fisheye lens in image viewing are suitable for certain applications{~\cite{rao1994table,yang2022pattern,elmqvist2008melange}}, but their distortion effects and extensive adaptation requirements limit their suitability in text-dense environments like computational notebooks.
Overview+detail design, featuring a separate overview window that displays a thumbnail of the entire content, aids users in identifying their position and finding specific sections. 
This approach is integrated into some text editors, IDEs, and even computational notebooks like Google Colab{~\cite{GoogleColab}}, where it presents an "outline" view based on the markdown hierarchies.
Nevertheless, users require extra and explicit effort to create and maintain the hierarchies or summaries.
In this research, we hypothesize that the inherent spatial and physical navigation offered by immersive environments can streamline notebook navigation.

Analyzing different hypotheses often mandates the development of multiple versions of analyses or implementations, followed by result comparisons.
Managing these versions poses challenges and is often error-prone. 
Hartmann et al.~\cite{hartmann2008design} developed interfaces presenting results from various alternatives in a unified view to simplify comparison.
Weinman et al.~\cite{weinman2021fork} adopted these concepts to computational notebooks, proposing the ability to create multiple non-linear execution paths. 
Here, users can ``fork'' content from a chosen cell, but the approach's limitation to full code duplication and lack of support for simultaneous paths may restrict its use and increase maintenance.
As an alternative, Harden et al. introduced a 2D computation notebook, giving users the flexibility to organize cells and results bidimensionally, facilitating easier side-by-side comparisons~\cite{harden2022exploring}.
Inspired by these precedents, we aim to empower analysts to replicate only essential content for creating comparisons and to position comparison results adjacently for efficiency.

\subsection{Immersive Analytics}
Immersive analytics represents an emerging research field exploring the integration of novel interaction and display technologies to enhance data analytics, particularly through the lens of VR/AR~\cite{marriott2018immersive,ens2021grand,zhao2022metaverse}.
Current studies in immersive analytics predominantly emphasize the data visualization aspect and have identified several key benefits of using VR/AR.
For instance, previous work reported rendering 3D network graphs in VR to be more effective than on flat screens due to the added dimension to declutter the visual information{~\cite{kwon_study_2016, cordeil2016immersive, yang_origin-destination_2019}}.
The large display space in VR also permits users to organize content spatially{~\cite{satriadi2020maps,hayatpur2020datahop,lin2021labeling}}, enhancing physical navigation such as walking and head movement, found to be more effective than virtual methods like pan\&zoom{~\cite{ball_move_2007}}. 
Research further highlights spatial memory plays a crucial role in VR/AR information retrieval~\cite{yang_virtual_2021}, and VR/AR's accurate motion tracking offers opportunities for intuitive interaction designs, like the Tilt Map{~\cite{yang2020tilt}} that enables switching between 2D and 3D visualizations.
It's worth noting that the cited examples merely offer a snapshot of the myriad VR/AR advantages documented in literature rather than an exhaustive list.

Building on empirical evidence that attests to the advantages of VR/AR in data visualization, researchers in the field of immersive analytics have begun to explore whether these benefits can be extended to the broader scope of data analytics.
For instance, In et al.~\cite{in2023table} developed a tool that facilitates gesture-based interactive data transformation within a VR environment.
In a similar vein, Lisle and Davidson et al.~\cite{lisle_evaluating_2020,lisle_sensemaking_2021,davidson2022exploring} introduced the concept of an ``Immersive Space to Think,'' leveraging the expansive display capabilities of VR/AR for improved text content management and insight generation. 
Luo et al.~\cite{luo2022should} examined strategies for spatially organizing documents in AR settings.
In alignment with these pioneering efforts, our research aims to explore the potential benefits that immersive environments could offer to computational notebook applications.

 \begin{figure*}
    \centering
    \includegraphics[width=1\textwidth]{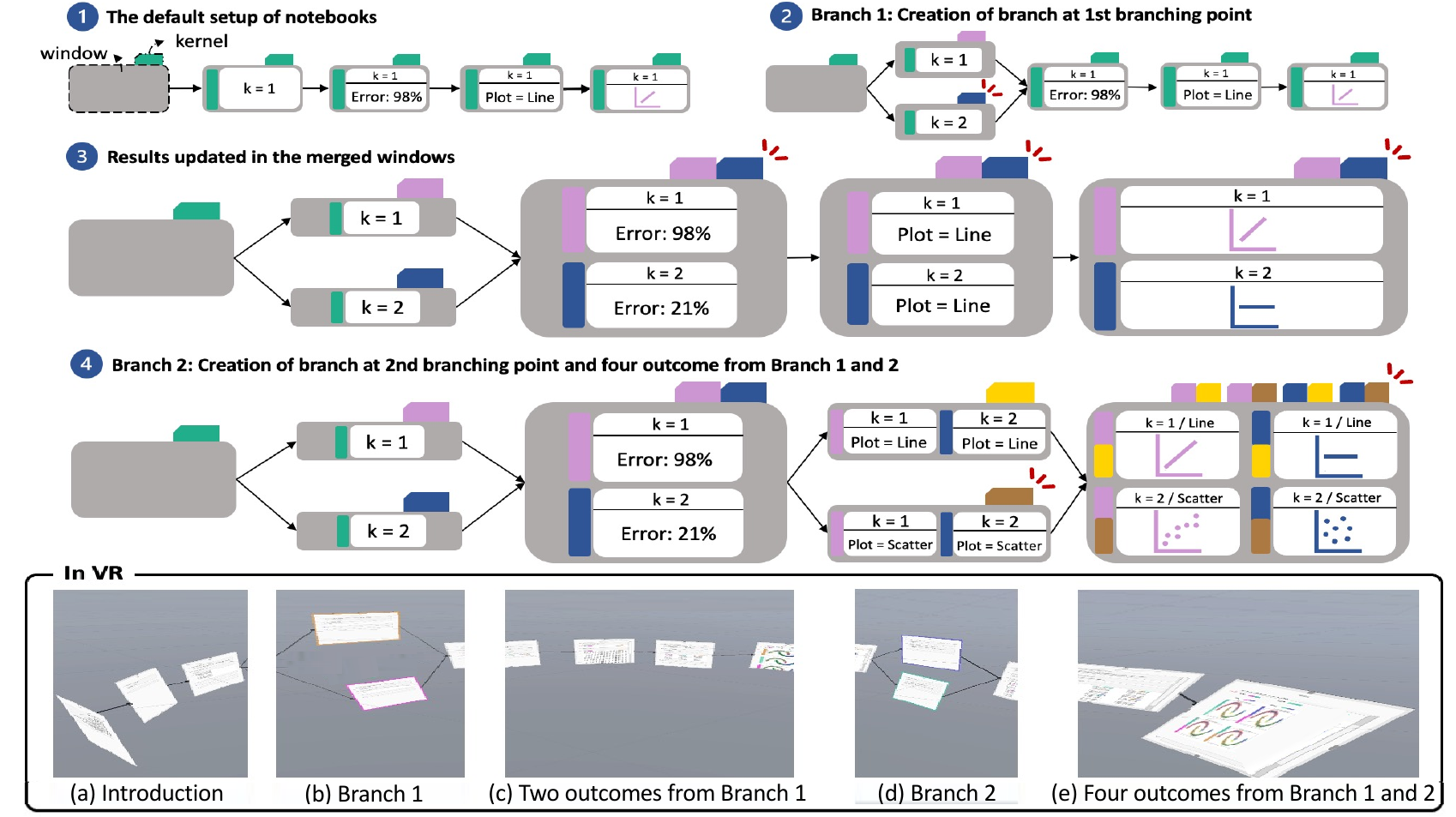}
    \caption{Visual representation (top) of the Branch\&Merge mechanism within computational notebooks, as instantiated in a VR (bottom). Different hues within the results are employed to illustrate the independent storage of variables across branches. The figure is segmented into five key operations: (a) regular notebooks, (b) initial creation of the branch, (c) merge back to a singular code execution path, (d) initiation of the second branch, and (e) final merging process to facilitate value comparisons across all initiated branches.
    }
    \Description{Demonstration of Branch\&Merge mechanism. The top part of the figure demonstrates the Branch\&Merge process in computational notebooks, presented through a series of steps in a gray, rounded rectangle window. Attached to this window is an angular rectangle symbolizing the kernel, which is the core system in each notebook. Within this kernel, different shades are used to represent the distinct storage of variables in various branches, along with their respective results. The bottom part depicts the same Branch\&Merge in a VR setting. Details are in the caption and Sec. 3.2 Embodied Branch\&Merge for Comparison.}
    \label{fig:branch_merge}
\end{figure*}

\section{Adapting Computational Notebooks to VR}
\label{sec:design_implementaion}

Our primary objective is to examine the user experience of computational notebooks in VR and to explore the potential advantages VR may offer. 
A crucial initial step is adapting the computational notebook system for VR use. While the desktop version of the computational notebook is well-established, transitioning it to a VR setting introduces unique challenges. 
These include determining how to visually represent and spatially position the notebook in the VR space and identifying necessary interactions to facilitate its use in this immersive environment.
As a preliminary step for designing VR-compatible computational notebooks, we aimed for a smooth transition for analysts by maintaining design consistency with familiar desktop counterparts, while also leveraging VR's distinctive capabilities where beneficial. 
This section details our design objectives, centered on enhancing the \textit{navigation} and \textit{comparison} functionalities of computational notebooks, and discusses our primary design considerations and decisions.

\subsection{Design Goals}
\textbf{Navigating} a computational notebook involves an analyst shifting their focus to a different section of the notebook by actively changing the content displayed in their field of view (FoV). 
Navigation is fundamental to various tasks performed by analysts using computational notebooks. 
Analysts frequently navigate between various parts of their analytical pipeline during exploratory data analysis to derive insights from data. 
Similarly, navigation is essential when taking over someone else's project to understand their process or when compiling a final report from the analysis conducted{~\cite{kang2021toonnote}}.
Building upon prior research~\cite{bowman_3d_2004,nilsson_natural_2018,plumlee_zooming_2002,plumlee_zooming_2006,yang2020embodied}, we define navigation as a multi-component process, primarily involving locating target sections and then moving towards them.
With a conventional desktop computational notebook, it is challenging for analysts to recall the location of off-screen targets and accurately navigate (``scroll'') to these locations.
Therefore, our goal \textit{is to enable analysts to quickly identify (locate) and reach (move to) their desired sections within the computational notebook in VR}.

\textbf{Comparing} outcomes derived from varying parameters or methods is a frequent task for analysts working with computational notebooks{~\cite{liu2019understanding}}. 
This process involves two stages: initially establishing the comparisons by coding tests for different parameters or methods, and subsequently examining the results to make informed choices regarding these parameters or methods.
In the setting of traditional desktop computational notebooks, analysts typically rely on their memory for comparison tasks, externalize results for comparisons, write specialized code to produce multifaceted results, or replicate the notebook for comparison purposes. 
However, each of these methods has its limitations, either placing a significant cognitive load on the analyst's working memory or complicating the management of code and content.
Consequently, our objective \textit{is to enable analysts to intuitively generate comparisons and easily review all generated results}.

\begin{figure*}
    \centering
    \includegraphics[width=0.9\textwidth]{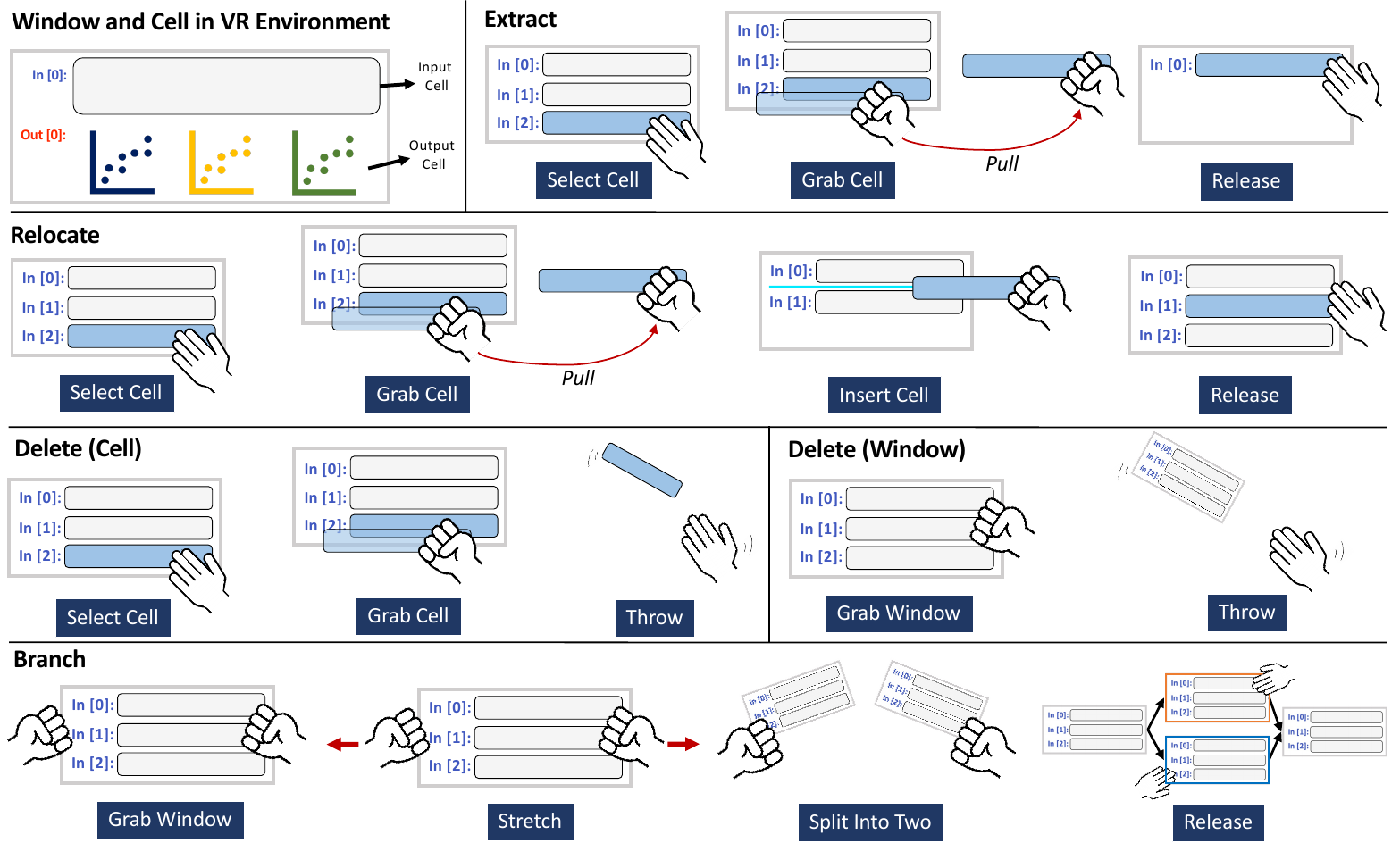}
    \caption{Illustrations of gestures interactions in the \VRT{} environment.}
    \Description{Required gestures to perform interaction in VR environments. Selected cells are marked in blue, whereas inactive ones appear in light grey. A light blue line serves as a guide, showing where to place a selected cell in a designated spot for "Relocate" interaction. Additionally, varying colors in the scatter plots at the upper left and lower right corners represent data originating from a variety of distinct hypotheses. Further details are in Sec. 3.2 Additional Notebook Interactions.}
    \label{fig:vr_gesture}
\end{figure*}

\subsection{Adaptations}
To investigate the advantages that VR could offer to computational notebooks, we designed and implemented adaptations that capitalize on VR's distinctive display and interaction features, with a focus on enhancing the \textit{navigation} and \textit{comparison} experiences.
We specifically focused on facilitating physical navigation within the VR environment and enhancing comparison tasks through an embodied branch\&merge gesture. Furthermore, we also developed other essential interactions tailored to the notebook's functionality in VR.

\textbf{Enabling Physical Navigation.} 
Physical navigation involves utilizing bodily movements, such as head rotation or walking, to explore an information space, like accessing various sections of a computational notebook. 
Research by Ball et al.~\cite{ball_move_2007} demonstrated that this form of navigation is more efficient than virtual navigation (pan and zoom) in the context of browsing geographic maps on large display walls. 
In VR, the ability to spatially arrange content offers a unique opportunity for physical navigation, which we anticipate could enhance notebook navigation compared to desktop versions. 
To establish an environment conducive to physical navigation, we implemented the following adaptations.

\textit{Adding one hierarchical layer.} Desktop computational notebooks employ a linear arrangement of cells and outputs within a singular window. Yet, transplanting this design directly to an immersive environment poses challenges. Extending the single window to accommodate all content could result in an impractically long display, potentially falling outside the user’s reach. To convert a notebook into a suitable format for spatial distribution, we introduce an additional hierarchical layer to the content layout: cells and outputs are organized within individual windows, and these windows are interlinked to compose a complete notebook, see \autoref{fig:teaser}.

\textit{Applying a curved layout.} 
Drawing on observations from Andrews et al.~\cite{andrews2010space} regarding the layout of multiple windows, we adopted a commonly used horizontal window placement strategy.
This linear arrangement, signified by directed arrows from left to right, not only clarifies the sequential order of windows but also ensures that all windows fall within the user's vertical reach. 
To optimize the curvature of this arrangement, we consulted Liu et al.'s findings~\cite{liu2020design,liu2022effects} for our initial layout placement, which indicate that a semi-circular layout generally surpasses both flat and full-circle configurations.

\textbf{Embodied Branch\&Merge for Comparison.} 
In hypothesis testing via comparisons, analysts frequently employ the strategy of creating additional copies and modifying relevant content, such as adjusting a variable's value or invoking a different function, as highlighted by Weinman et al.~\cite{weinman2021fork}.
In response, they introduced an interactive tool named ``fork it,'' enabling users to create a concurrent copy with a button click.
Adapting this interactive concept to immersive settings, we designed an embodied gesture for duplication:
users grasp the window they wish to replicate with both hands and then stretch it until a specific threshold, as illustrated in \autoref{fig:vr_gesture}---Branch.
The user can freely place the newly created windows in space.

Our enhanced hierarchical structure in notebook content organization offers increased flexibility for hypothesis evaluation through non-linear branching. 
This enables precise content duplication at the window's granularity, as opposed to copying subsequent content, as seen in the ``fork it''~\cite{weinman2021fork}.
For instance, when an analyst intends to probe different predefined cluster values in K-means clustering, they can produce branches for various cluster assignments. 
The subsequent visualization code used to assess clustering results remains unduplicated.
Moreover, our system supports branching at multiple points simultaneously, unlike prior systems that were limited to single-point branching. 
For instance, consider having a linear notebook without any branch initially, illustrated in Fig.~\ref{fig:branch_merge}-\one{}.
Subsequently, the user can create a branch at any point, as demonstrated in  Fig.~\ref{fig:branch_merge}-\two{}. 
The creation of a branch leads to added results for comparisons in all the subsequent windows. 
The number of results is equal to the number of branched windows, say two, as depicted in Fig.~\ref{fig:branch_merge}-\three{}. 
Furthermore, the analyst can create another branch as demonstrated in  Fig.~\ref{fig:branch_merge}-\four{}, and the subsequent window (only one last window in this case) will have four results produced by all combinations of the two branches.

\textbf{Additional Notebook Interactions.}
Computational notebooks come with fundamental interactions for content management, such as creating, deleting, and moving cells, typically one cell at a time. 
Initially, we explored a toolbar-based design situated at the top of the window, mirroring the traditional desktop-based computational notebook interface. 
Nevertheless, our internal evaluations highlighted challenges with using a pointer for button interactions, confirming findings from previous studies~\cite{nandi2013gestural, rzeszotarski2014kinetica}.
A recent study by In et al. highlighted the advantages of gesture-based interactions over the window-icon-menu-pointer (WIMP) design~\cite{in2023table} in immersive settings.
Consequently, we pivoted to designing intuitive gesture interactions for the immersive environment, which are also demonstrated in \autoref{fig:vr_gesture}.

\begin{itemize}
    \item \textit{Extract}: Users can deploy a ``grab \& pull'' gesture to detach a cell from its window, creating a new window for the selected cell. When extracting multiple cells, users can first select them and then employ the ``grab \& pull'' gesture.
    
    \item \textit{Delete}: To remove a cell, users can utilize the ``grab \& throw away'' gesture. 
    For multiple cells, after selection, the same gesture is employed. Entire windows can similarly be discarded with this gesture.
    
    \item \textit{Relocate}:  Users can ``grab'' a cell and ``drag'' it to a new position, whether within its initial window or to a different one. To relocate multiple cells, users should first select them and then execute the ``grab \& drag'' action. 
\end{itemize}

Additional interactions include ``grab \& move'' for repositioning windows and "pinch-to-zoom" for resizing, where users collide their hands with a window and move them apart to enlarge or together to minimize. 
Beyond content and view adjustments, a single button on the user's left hand, inspired by Yang et al.~\cite{yang2020embodied}, proved more efficient than attaching buttons to each window. 
This ``Run'' button executes the selected cell and all subsequent cells.


\section{User Study and Evaluation}
\label{sec:study}
To systematically answer the research question, we designed and conducted a controlled user study.
Primarily, our objective was to explore the potential advantages of utilizing an immersive environment. 
To this end, we compared our immersive computational notebook implementation with its \DesktopT{} counterpart. 
Additionally, we sought to enrich empirical evidence supporting the merits of the ``branch\&merge.''
In summary, our study encompassed four conditions: \desktopLinearT{}, \desktopBranchT{}, \vrLinearT{}, and \vrBranchT{}. 
These conditions enabled a systematic investigation of the effect of two variables: the computing environments (\fDesktop{}\BreakableChar{} vs. \fVR{}) and the comparison techniques (\fLinear{}\BreakableChar{} vs. \fBranch{}).

\subsection{Study Conditions}
\label{sec:conditions}

Our \fBranch{} design in the immersive computational notebook (\vrBranchT{}) is presented at \autoref{sec:design_implementaion}---Branch\&Merge. 
Meanwhile, our \desktopBranchT{} implementation shared a similar idea but utilized a button to duplicate a window instead of using an embodied gesture. 
On the other hand, the creation of branches was not permitted in both the \desktopLinearT{} and \vrLinearT{} conditions.
In the following, we detail our \fDesktop{} and \fVR{} implementations.


\textbf{\fDesktopBold{}}:
In the \DesktopT{} environment, our design emulates the features and functionalities typical of standard computational notebooks, where interactions are facilitated via a mouse and keyboard.
To navigate different segments of the notebook, we incorporated vertical scrolling, a standard navigation method in many \DesktopT{} applications. 
Users can navigate by using the mouse scroll or dragging the scrollbar. 
For other content interactions, we also follow the standard computational notebook designs: users use the mouse to select the target and click the buttons to execute specific commands, like extract, delete, and relocate.
To ensure an equitable comparison and to control for potential confounders, we integrated features detailed in \autoref{sec:design_implementaion}. 

\textbf{\fVRBold}:
We detail our primary computational notebook designs and implementations for \VRT{} in \autoref{sec:design_implementaion}.
This section focuses on vital design choices not strictly tied to computational notebook features. We opted for bare-hand interaction over using controllers, aiming for a more intuitive and immersive user experience. 

However, text input remains a crucial aspect of immersive computational notebooks. Meta has pioneered a technology that brings physical keyboard tracking into \VRT{}~\cite{MetaKeyboard}, which we initially adopted.
However, Meta's design primarily suits a seated work environment, as carrying and typing on a physical keyboard while navigating in \VRT{} is impractical.
While Davidson et al.~\cite{davidson2022exploring} proposed using a mobile table for the physical keyboard, their approach inadvertently tethered users to the table, reducing spatial exploration. 
To encourage fuller utilization of physical navigation in \VRT{}, we decided not to adopt their method, opting for the standard virtual keyboard method instead.
Our first virtual keyboard implementation employed an ``on-demand'' approach, appearing during text interaction and vanishing when interacting with other elements.
However, this approach was changed due to internal tests showing it often activated by mistake, leading to cluttering the interface. 
The revised design, inspired by Yang et al.~\cite{yang2020embodied}, attaches the keyboard to the left palm, with improvements to prevent unintentional activations by requiring users to intentionally look at their palm to activate it.

\begin{figure}
    \centering
    \includegraphics[width=1\columnwidth]{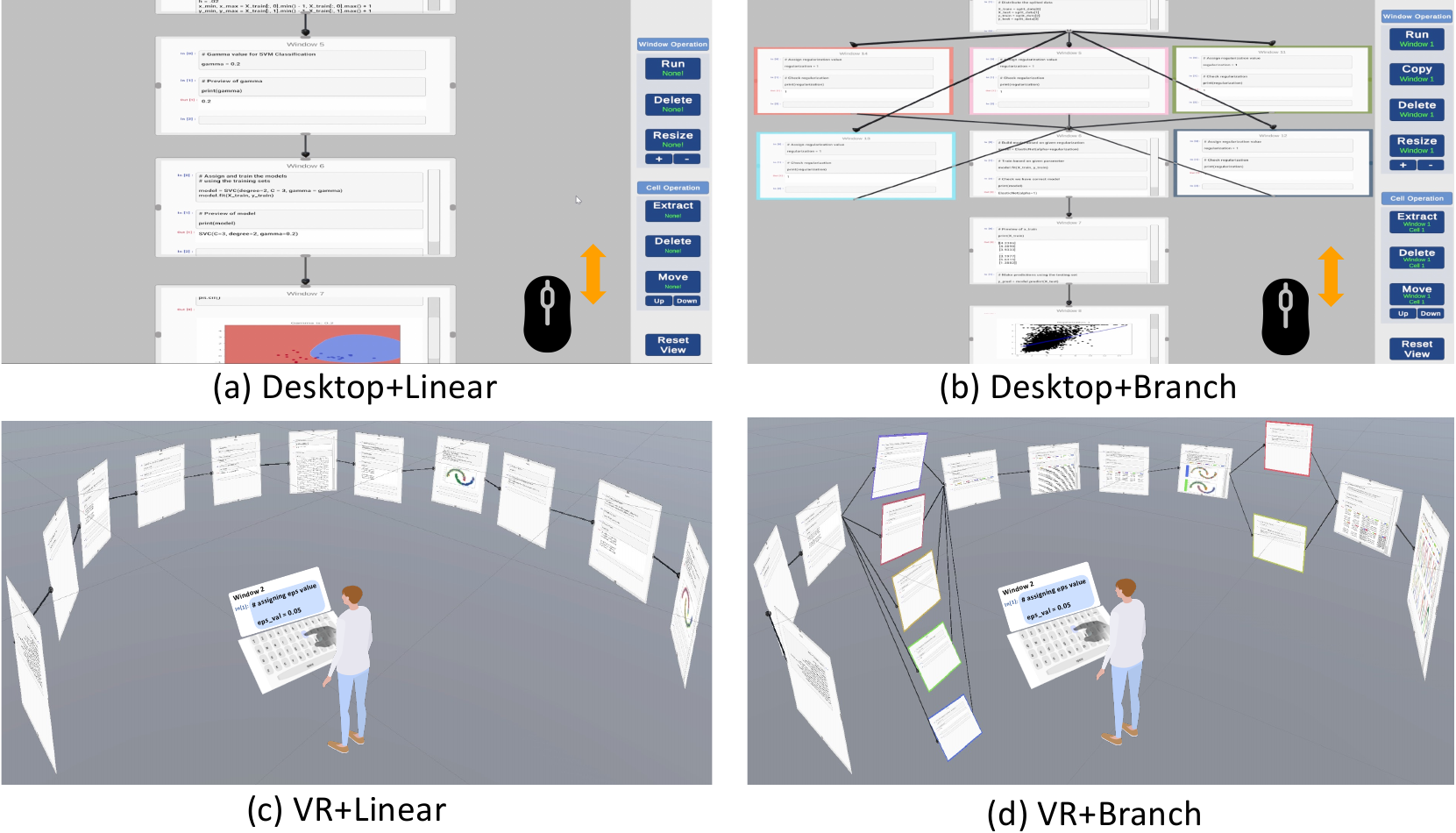}
    \caption{Demonstrations of conditions tested in our study. User scrolling to navigate in \DesktopT{} (top), and walking and rotating their body to navigate in \VRT{} (bottom). In addition, the user used a physical keyboard to write codes in \DesktopT{}, while used a virtual keyboard in \VRT{}.}
    \Description{Four conditions that were used in the study. Details are in the Sec 4.1 and 4.3.}
    \label{fig:all_condition}
\end{figure}

In \VRT{} settings, interacting with distant windows presents a challenge for various text interactions.
We selected a font size that ensured text readability from the initial viewing distance, eliminating the need for users to continuously adjust their proximity for code visibility.
While the text remains readable from afar, pinpointing and selecting a precise point within the text, such as a specific entry, remains challenging.
To facilitate precise interaction, we implemented a design inspired by Voodoo dolls~\cite{pierce1999voodoo}, where a proximate copy of the interacting cell is generated for the user. 
Interactions between this close copy and the original notebook window are synchronized, as illustrated in Fig. \ref{fig:all_condition} (c) and (d).

\subsection{Task and Data}
\label{sec:task_and_data}

Computational notebooks are versatile tools that facilitate a wide range of analytical tasks, from creating new notebooks to utilizing existing code for data analysis. 
However, requiring participants to extensively write code from scratch in a controlled study could be time-consuming and introduce confounding factors outside the scope of our investigation. 
Meanwhile, it's increasingly common for analysts to revisit or leverage existing code, whether it's from inheriting someone else's project or reusing their previous work~\cite{alspaugh2018futzing}. 
To maintain the focus of our study on the aspects of navigation and comparison, we simulated a scenario where participants were provided with pre-existing code.

We structured two tasks to evaluate the navigation and comparison performance within our test conditions. 
To mitigate any learning effects, each participant was presented with four analytical methods, assigning one method to each test condition: K-Nearest Neighbors (KNN), Support Vector Machine (SVM), Density-Based Spatial Clustering of Applications with Noise (DBSCAN), and Linear Regression.
As mentioned, the order was determined by a balanced Latin square matrix.
The computational notebooks used in the study were sourced from the official scikit-learn documentation~\cite{scikit-learn}. 
These notebooks were standardized to approximately ten windows each, based on internal testing that showed our chosen length allowed for practical task completion times.
Moreover, the notebooks exhibited a consistent logical structure: they initiated with data loading, data transformation, modeling, and concluding with the visualization of model outcomes.
Details of the specific study tasks, formulated within this overarching context, are presented subsequently. 
Supplementary materials containing all study stimuli are provided.

\textbf{Task 1. Navigation.}
To assess the performance across different types of navigation, we designed tasks that involved both single-stop and multi-stop (specifically, two-stop) navigation scenarios.
The single-stop task had participants identify and delete an error-causing cell called \textit{deletion}, while the two-stop task involved identifying and correctly repositioning a misplaced cell called \textit{relocation}.
Participants could identify these target cells by inspecting the cells' output; notably, cells generating errors and their subsequent cells would not yield output.
This task was designed to be navigation-intensive, enabling us to extract nuanced differences in performance and user experience related to navigation activities. 
Consequently, our primary aim was to investigate the impact of computational environments---\VRT{} vs. \DesktopT{}---on navigation efficiency and user experience.

\textbf{Task 2. Comparison.} 
Following Task 1, we introduced Task 2, designed to simulate real-world hypothesis testing.
Participants evaluated two parameters within various analytical methods, such as cluster numbers and distance metrics in the KNN method, to determine the optimal parameter combinations based on the visualized results. 
To ensure consistent experimental conditions, we standardized the spatial distance of relevant windows (i.e., the two windows containing ``what-if'' tests and the result window) across all trials.
Minimal text input was required from participants due to pre-commented code, simplifying the process of parameter adjustment. 
Although our primary focus was on comparison, navigation was inherently involved in task completion.

\subsection{Experimental Setup}
In the \VRT{} configuration, we utilized the Meta Quest Pro headset, providing a resolution of $1800\times1920$.
For \DesktopT{} conditions, a standard 27'' monitor was deployed, featuring a 2560x1440 resolution.
The Meta Air Link feature was used for the \VRT{} environments, enabling a tether-free experience by leveraging the PC for computations while the headset managed to render. 
This configuration allowed participants to freely navigate the $16 m^2$ space without worrying about cable impediments. 
Within the \VRT{} setting, participants started at the center of the space, presented with ten notebook windows, each measuring $0.35x0.30 m^2$, arranged semi-circularly at a distance of $1m$.
Conversely, \DesktopT{} participants sat at a desk, encountering initial notebook windows followed a linear structure and sized at on average $2000\times600$ pixels each.
Specifically, the notebook windows were displayed on the center of the monitor, allowing participants to view two or three notebook windows at the same time. 
Additionally, for the \desktopBranchT{}, we left empty spaces on both the left and right sides of the notebook windows to ensure sufficient space for branching.
This arrangement was mirrored in the \desktopLinearT{} to ensure consistency across the study settings, as shown in \autoref{fig:all_condition} (a) and (b).

\subsection{Participants}
We recruited 20 participants (16 male, four female, ages 18 to 35) from a university mailing list. 
The recruitment was based on their knowledge of data science and machine learning algorithms, which was screened using an eight-question quiz (provided in the supplementary material). 
Participants need to answer six out of eight to be eligible for the study.
Out of the 22 respondents, 20 were invited to participate in the study based on the eligibility requirement.
Regarding \VRT{} experience, seven of the participants use \VRT{} weekly, and the remaining thirteen have no prior \VRT{} experience. 
All participants had either normal vision or vision corrected to normal.
For their time and contribution to the study, each participant was compensated with a \$20 Amazon Gift Card.

\subsection{Design and Procedures}
Our user study followed a full-factorial within-subjects design, with conditions balanced using a Latin square (4 groups).
The study, on average, took less than two hours.
Participants were initially welcomed and reviewed a consent form.
Then, we briefly introduced the study's objectives and procedural steps.
Following this introduction, participants proceeded to the various components of the study as follows:

\textbf{Preparation}:
We asked participants to adjust the chair height to a comfortable level for the \DesktopT{} condition and adjust the Quest Pro headset for the \VRT{} condition before they started the training session. 
We confirmed that all participants were in comfortable conditions and could see the text in all display environments clearly. 

\textbf{Training}:
We initiated our study by standardizing computational notebook terminologies, recognizing the potential for varied interpretations. 
The training was provided only when participants first encountered a computing environment (i.e., \fDesktop \BreakableChar{} or \fVR). 
This was due to the consistent operational logic within each environment, with differences only in the comparison task. 
In the training, participants viewed operational demonstration videos. 
Post-viewing, we verified their understanding, asking them to replicate study tasks using a different algorithm, k-mean. 
Participants were free to inquire about operations or tasks. 
The training sessions, particularly for the VR condition, were extended to give participants enough time to become comfortable with the immersive environment. 
This approach was adopted to address potential VR-related issues such as discomfort, learning effects, and novelty bias.
In summary, the training was completed once participants achieved proficiency in tasks and operations, which generally took 10-15 minutes.


\textbf{Study Task}:
Upon completion of the training session, participants proceeded to the study task. 
To ensure they had enough understanding of what would be expected, we provided comprehensive context, including a brief explanation of the algorithms and the tasks they needed to complete.
Participants had no time limit for task completion but were encouraged to prioritize accuracy and efficiency.
For the \VRT{} environment, we reset the participants' position to the center of the room and had them face the same initial direction before each study task started.

\textbf{Questionnaires}.
\textit{Post-Condition Questionnaires}: upon completion of each condition, participants were required to fill out a Likert-scale survey. 
This was adapted from the System Usability Scale (SUS) and NASA Task Load Index (TLX) to record their subjective experiences. 
Additionally, they were asked to provide qualitative feedback concerning the pros and cons of the condition they had just interacted with. 
\textit{Post-Study Questionnaires}: Once all the study tasks and post-condition questionnaires were completed, participants were asked to rank the study conditions based on their overall experience.

\subsection{Hypotheses} 
\label{sec:hypo}
We aimed to validate whether our designs and implementations met our established design goals. Consequently, we formulated hypotheses grounded in empirical findings from prior research and the testing conditions described in \autoref{sec:conditions}.

\textbf{Navigation in \VRT{} and on \DesktopT{} ($H_{nav}$).}
We hypothesized \VRT{} would provide faster navigation than \DesktopT{}.
\VRT{} offers a large display space to lay out an entire notebook, allowing participants to navigate by physically walking in the space or rotating their heads.
In contrast, \DesktopT{} presents only part of a notebook at a time, requiring participants to scroll up and down for navigation.
Previous studies indicate that physical navigation---employed in our \VRT{} conditions---is more effective than virtual navigation, as used in \DesktopT{} conditions~\cite{ball_move_2007}.
Furthermore, \VRT{} has been demonstrated to enhance spatial awareness, thereby aiding in the recall process during multi-stop navigations~\cite{krokos2019virtual,yang_virtual_2021}.
We believe these established advantages of \VRT{} are generalizable to computational notebooks.

\textbf{Comparison with and without \BranchT{} ($H_{comp-branch}$).}
We expected that the incorporation of the \BranchT{} feature would facilitate comparison tasks.
Earlier studies have reported favorable user experiences with similar functionalities in computational notebooks~\cite{weinman2021fork,harden2022exploring}.
Building on these insights, we introduced additional features, such as merging post-branching, to minimize visual clutter and spatially organize results. 
We aim to provide quantitative empirical data to highlight the effectiveness of the \BranchT{} functionality.

\textbf{Comparison in \VRT{} and on \DesktopT{} ($H_{comp-env}$).}
We anticipated that \VRT{} could outperform \DesktopT{} in performing comparisons.
In the \LinearT{} conditions of \DesktopT{} and \VRT{}, the comparison would be intrinsically linked to the navigation efficiency, as participants would only view one result at a time in both conditions.
Given this, the navigational advantages of \VRT{} are expected to positively impact comparison tasks in \LinearT{} conditions. 
For the \BranchT{} conditions, we considered our designed \VRT{} embodied gesture for branch creation would be intuitive, thereby facilitating the process.
Moreover, the large display space in \VRT{} could allow participants to view all results simultaneously, expediting the visual assessment process.

\subsection{Measures}
In our study, we gathered quantitative data to evaluate our hypotheses.
For the navigation task, we logged the \textit{time} taken by participants to complete one-stop (deletion) and two-stop (relocation) navigations under each condition. 
\textit{Completion times} for the comparison task were also recorded, as was the frequency of ``Run'' button presses, indicating execution. 
For potential subsequent analyses, we also logged user interaction and tracked objects in the scene.

After participants engaged with a specific condition, they were asked to complete a survey using a 7-point Likert scale to gauge their perceived \textit{physical} and \textit{mental demands}, \textit{engagement}, and the \textit{effectiveness} of that condition. 
To further explore the nuances of each condition, semi-structured interviews were conducted, highlighting both strengths and areas for improvement for each condition. 
Concluding the study, participants provided an overall ranking of their user experience within the testing conditions.



\section{Results}

\begin{figure}
    \centering
    \includegraphics[width=0.65\columnwidth]{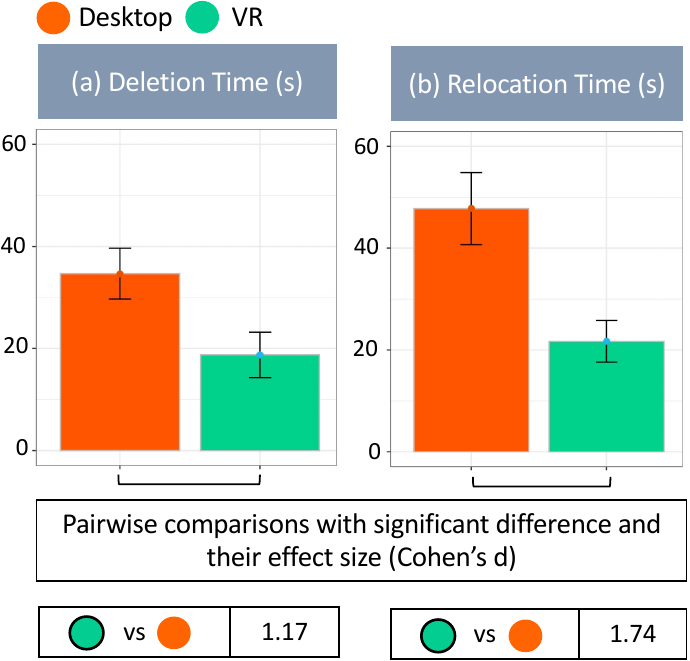}
    \caption{Completion time for \VRT{} and \DesktopT{} in the navigation task. (a) the time spent for completing deletion, and (b) the time spent completing relocation. Solid lines indicate statistical significance with $p < 0.05$. The tables below show the Cohen's D effect sizes for significant comparisons. Circles with black borders indicate the condition with better results.}
    \Description{The bar charts display the quantitative data for deletion time and relocation time, measured in seconds, for Task 1. These charts include error bars representing 95\% confidence intervals. Details can be found in Section 5.1.}
    \label{fig:quant_findings_navigation}
\end{figure}

\begin{figure*}
    \centering
    \includegraphics[width=0.9\textwidth]{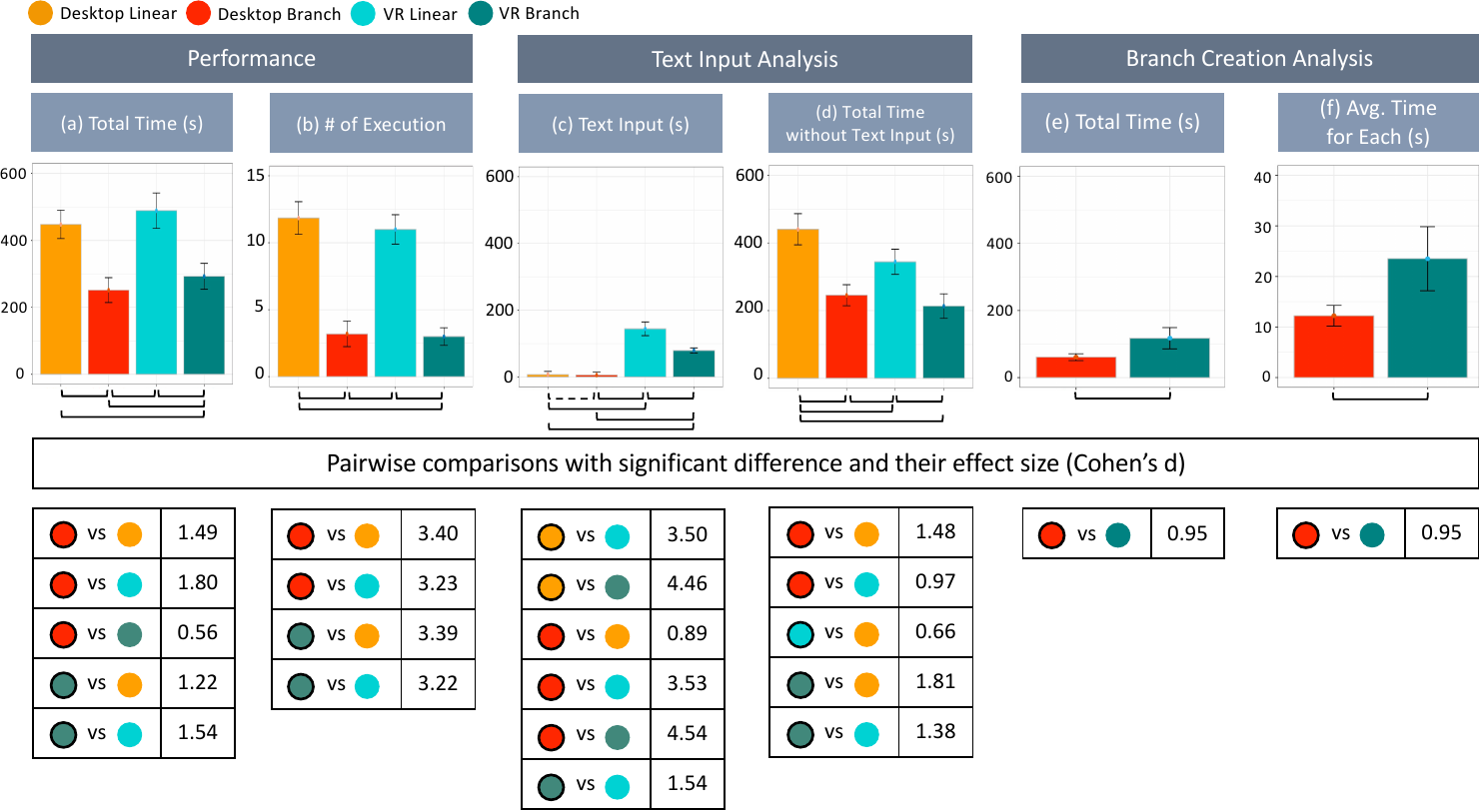}
    \caption{Analysis of four testing conditions in the comparison task. (a) the time spent completing the task, (b) the number of executions performed, (c) the time spent for text input, (d) the time spent completing the task excluding the text input interaction time, (e) the total time spent for creating all branches in \DesktopT{} and \VRT{}, (f) the average time spent for creating a branch in \DesktopT{} and \VRT{}. 
    Solid lines indicate statistical significance with $p < 0.05$, and dashed lines indicate $0.05 < p < 0.1$.
    The tables below show the Cohen's D effect sizes for significant comparisons. Circles with black borders indicate the condition with better results.}
    \Description{The bar charts display the quantitative data for several aspects of Task 2, including total time taken, time dedicated to text input, time spent excluding text input, time used for creating a branch, and the average time required to create each branch, all quantified in seconds. Additionally, the charts show the number of codes executed during the task. Error bars indicating 95\% confidence intervals are also included in these charts. For more detailed information, refer to Section 5.1.}
    \label{fig:quant_findings_comparison}
\end{figure*}

\label{sec:results}
In this section, we present the statistical analysis of our collected data, outline the strategies participants employed to manage the display space, and provide summarized qualitative feedback for each condition.
We documented significance at levels of $p < 0.001 (***)$, $p < 0.01 (**)$, $p < 0.05 (*)$, and $p < 0.1 (\cdot)$.
Additionally, we present mean metrics with a 95\% confidence interval (CI) and use Cohen's d to determine the effect sizes of significant differences. 
Comprehensive statistical analysis results can be found in the supplementary materials.

\subsection{Quantitative Results}
\label{sec:results-quan}

All participants successfully completed the study tasks, resulting in no variance in accuracy metrics. In the rest of this section, we present results from other measurements that highlight performance differences among the conditions.

\textbf{Completion time for the navigation task.}
The computing environment significantly influenced the time taken for deletion (one-stop navigation) and relocation (two-stop navigation), with both $***$.
On average, \VRT{} only required nearly half the time compared to \DesktopT{}, exhibiting statistical significance and large effect sizes.
To be more specific, \DesktopT{} witnessed a considerable 38.0\% increase in completion time (from avg. 34.7s to avg. 47.8s) between deletion and relocation, \VRT{} exhibited only a 15.7\% increase (from avg. 18.8s to avg. 21.7s) (refers to \autoref{fig:quant_findings_navigation}).
Consequently, we accept $H_{nav}$.

\textbf{Completion time for the comparison task.}
We found that having the branching feature had a significant effect on time and execution number ($***$), and the interaction of the computing environment and having the branching feature also had a significant effect on time and execution number ($***$).
We did not find the computing environment had a significant effect.

\BranchT{} conditions were significantly faster than \LinearT{} conditions, for both \VRT{} and \DesktopT{} ($***$), with large effect sizes, see \autoref{fig:quant_findings_comparison} (a). 
Additionally, having \BranchT{} also significantly reduced the number of executions required for both \VRT{} and \DesktopT{}, with large effect sizes, see \autoref{fig:quant_findings_comparison} (b). 
Thus, we accept $H_{comp-branch}$.
On the other hand, we did not observe \VRT{} outperforming \DesktopT{} in the comparison task.
In fact, \vrBranchT{} took longer than \desktopBranchT{}.
Therefore, $H_{comp-env}$ cannot be accepted.

\textbf{Text input time analysis.}
In our observations, we noted that participants devoted a considerable amount of time to text input within the VR environment, despite our efforts to streamline and enhance the text input experience, as outlined in \autoref{sec:design_implementaion}.
To systematically understand this influence, we analyzed the duration dedicated to text input across all test conditions. 
We discovered that the \VRT{} conditions necessitated significantly more time for text input compared to the \DesktopT{} conditions ($***$), with \vrLinearT{} taking more time than \vrBranchT{} ($***$)---both findings having large effect sizes. 
On average, text input consumed 144.3s, or 29.5\% of the total time, for \vrLinearT{}, and 79.3s, or 27.1\% for \vrBranchT{}. In contrast, desktop scenarios required a mere 7.2s (1.6\%) for \desktopLinearT{} and 5.9s (2.3\%) for \desktopBranchT{}, as shown in \autoref{fig:quant_findings_comparison} (c).
This data validates our observations, highlighting that text input in \VRT{} significantly hindered performance.

In a subsequent post hoc analysis, we excluded text input durations from all conditions and re-conducted the analysis, refer to \autoref{fig:quant_findings_comparison} (d). 
The results suggested that \vrLinearT{} was faster than \desktopLinearT{}, with a medium effect size and statistical significance ($**$). 
Although there appeared to be a trend with \vrBranchT{} outpacing \desktopBranchT{}, this distinction was not statistically significant.

\textbf{Branch creation efficiency.}
To further investigate the performance differences between \desktopBranchT{} and \vrBranchT{}, we examined the time taken for branch creation in both \DesktopT{} and \VRT{} environments, where we consider the total time of creating and placing branch windows, as illustrated in \autoref{fig:quant_findings_comparison} (e, f). 
This analysis encompassed the duration required for both the creation and positioning of the newly created branches. 
Our findings indicate that branch creation in \vrBranchT{} was notably slower than in \desktopBranchT{} ($***$), exhibiting a large effect size.

\begin{figure}
    \centering
    \includegraphics[width=0.9\columnwidth]{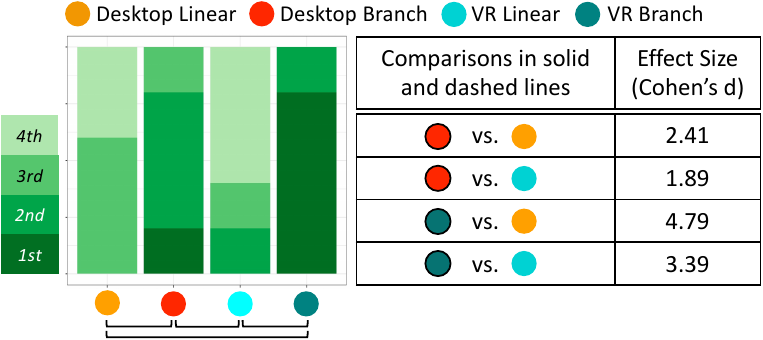}
    \caption{User ranking of overall user experience. Solid lines indicate significant differences with $p < 0.05$. The tables on the right show the Cohen's D effect sizes for significant comparisons. Circles with black borders indicate the condition with better results.}
    \Description{Stacked bar charts of user responses in their preference for the four experimental conditions. The color green is used to denote positive effects or favorable experiences. VR + Gesture was the most preferred condition to perform a given task. Details are in Sec 5.1.}
    \label{fig:quant_findings_comparison}
    \label{fig:ranking}
\end{figure}

\textbf{Ranking and ratings.}
Participants significantly favored the \BranchT{} conditions over the \LinearT{} conditions ($***$), as depicted in \autoref{fig:ranking}. 
Among them, \vrBranchT{} was most frequently cited as offering the best overall experience, with 80\% of participants placing it first. 
In contrast, 20\% of participants ranked \desktopBranchT{} as their top preference, and no participants considered \LinearT{} conditions as their most preferred condition.

Significant effects were also observed across all perceived metrics: \textit{physical demand} ($***$), \textit{mental demand} ($***$), \textit{engagement} ($***$), and \textit{effectiveness} ($***$).
\BranchT{} conditions outperformed \LinearT{} conditions in terms of mental demand ($***$), engagement ($***$), and effectiveness ($***$), all exhibiting large effect sizes. 
Moreover, \VRT{} conditions were found to be more physically demanding than \DesktopT{} conditions ($***$). 
\vrBranchT{} emerged as the most engaging condition, boasting an average rating of 6.65 out of 7 with a confidence interval of 0.46.
Impressively, 85\% of participants awarded it the highest engagement score.

\subsection{Layout Strategies}
\label{sec:results-layout}
To better understand how participants utilized the display space, we analyzed the final layouts of each trial to understand participant strategies in Desktop and VR settings, with detailed collections available in the supplementary materials.

Within the \LinearT{} conditions, we observed a limited number of layout-related interactions.
Notably, in \desktopLinearT{}, the vast majority of participants refrained from repositioning or resizing windows. 
Meanwhile, in \vrLinearT{}~, certain alterations to the initial layout were made: eight participants expanded the last result window, while two relocated select ``key'' windows---namely, those containing cells crucial for generating comparisons, and the result window itself.

\begin{figure}
    \centering
    \includegraphics[width=1\columnwidth]{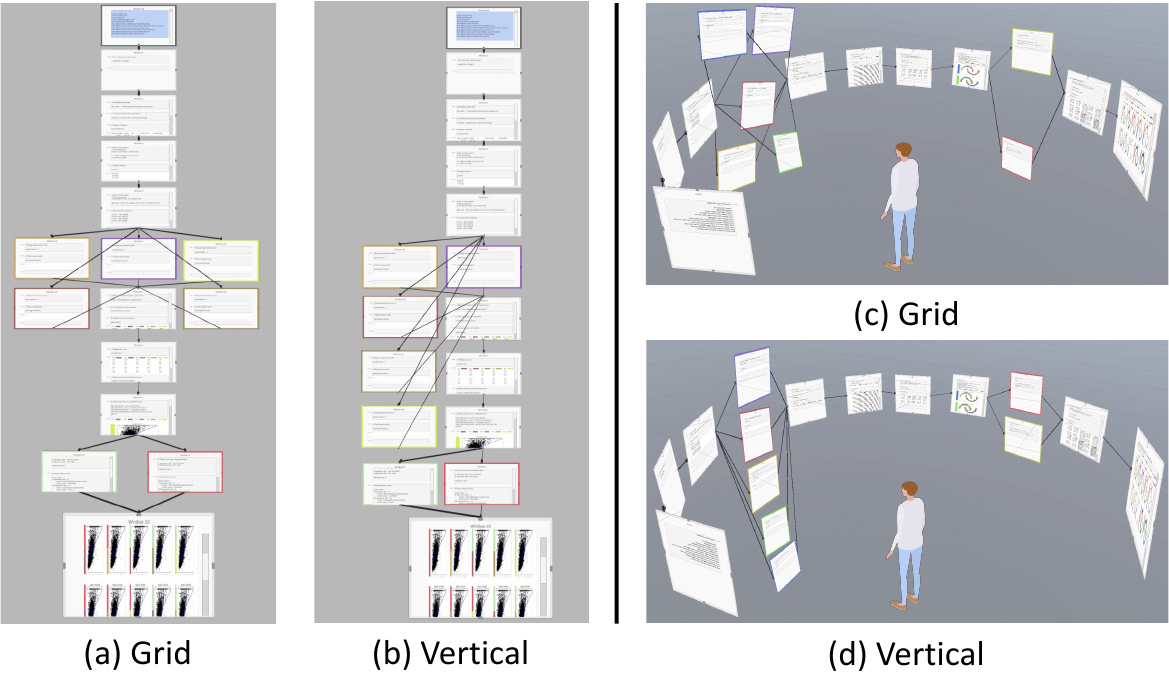}
    \caption{Two layout strategies used by our participants in \desktopBranchT (Left) and in \vrBranchT (Right).}
    \Description{Two layout strategies were used in desktop and VR conditions. Details are in Section 5.2.}
    \label{fig:organize_both}
\end{figure}

In contrast, the \BranchT{} conditions manifested a wider array of layout strategies, especially in terms of the placement of newly created branching windows. Within the \desktopBranchT{} condition, a dominant group of participants (15) arranged branching windows in a grid pattern. 
Initially, many aimed for a horizontal alignment, but due to spatial constraints, opted for additional rows, as depicted in \autoref{fig:organize_both} (a).
The other five participants chose a vertical layout, establishing a secondary column, as illustrated in \autoref{fig:organize_both} (b).
In \vrBranchT{}, similar to \vrLinearT{}, numerous participants enlarged and/or relocated particular ``key'' windows.
Within the large display space in \VRT{}, positioning of all branching windows could be orthogonal to the initial setup direction, eliminating overlap between connecting lines and windows---a choice made by 13 participants, showcased in \autoref{fig:organize_both} (c).
Meanwhile, seven participants opted to conserve vertical space, forming an extra column within a grid layout, displayed in \autoref{fig:organize_both} (d).

\begin{figure}
    \centering
    \includegraphics[width=1\columnwidth]{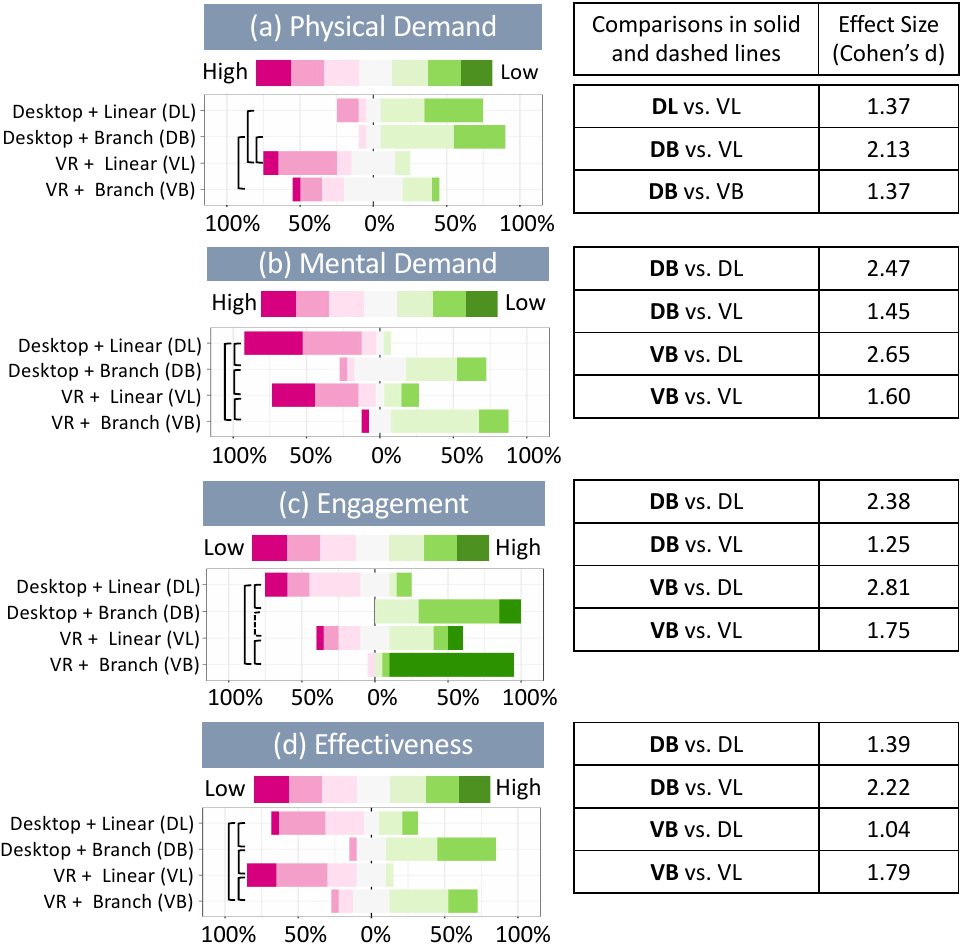}
    \caption{Subjective ratings on (a) physical demand, (b) mental demand, (c) engagement, and (d) effectiveness by task. Towards the right end of subfigures means better-perceived results. Solid lines indicate statistical significance with $p < 0.05$, and dashed lines indicate $0.05 < p < 0.1$.}
    \Description{Stacked bar charts showing the subjective ratings of physical demand, mental demand, engagement, and effectiveness among four experimental conditions. The color green is used to denote positive effects, while the color pink is used to denote negative effects. Details are in 5.3.}
    \label{fig:quant_findings_comparison}
    \label{fig:ratings}
    \vspace{-1.0mm}
\end{figure}

\subsection{Qualitative Feedback}
We conducted a qualitative analysis to identify common themes in user feedback for each condition. 
To systematically interpret our collected feedback data, two authors formulated a coding scheme rooted in the first five participants' feedback, which was subsequently applied uniformly to subsequent participants.
The top three codes and those mentioned by participants more than five times were reported for every condition (with frequency shown in parentheses). 
We concluded by summarizing the overarching insights gleaned from all the conditions. 
Comprehensive coding results are available in the supplemental materials.

\fdesktopLinearBold{} described favorable opinions with distribute cells into \textit{multiple windows are better than a single document} (7) and \textit{familiar} to typical computational notebook applications (4). 
However, drawbacks were noted, with concerns such as the need to \textit{scroll a lot} (14), \textit{text and target objects are hard to find} (8). 
These issues lead to \textit{inhibit task performance} (5).

\fdesktopBranchBold{} was considered positively, with \textit{easy to compare} the results of different parameter values (18), \textit{scroll less} (6), and \textit{fast} to perform comparisons (6). 
With the integration of the branch and merge feature, participants found it to \textit{lower the mental demand} and serve as an \textit{effective fine-tuning} method, according to three users.
However, some concerns were also noted, including that \textit{the initial linear layout is not effective} (9), and \textit{target objects were hard to find} (6).

\fvrLinearBold{} was praised for its inclusion of gesture interaction and physical navigation, with descriptions such as \textit{intuitive} (12), \textit{easy navigation} (9), 
and more \textit{fast} and \textit{effective} than WIMP (7). 
Participants also found it \textit{easy to understand the code} (7).
On the other hand, major issues were related to the \textit{text input difficulty} (14), with users stating that the provided virtual keyboard was challenging to interact with. 
 
\fvrBranchBold{}, similar to \vrLinearT{}, was considered as \textit{intuitive} (11), characterized by an \textit{effective initial layout} (11), and \textit{easy to understand the code} (4). 
It also demonstrated that the system provided a \textit{easy to compare} (9), \textit{easy navigation} (9), and was \textit{fast} to identify the optimal result (8). 
Unlike \desktopBranchT{}, participants in this condition described it as \textit{easy to organize} (8), with a \textit{large display place} (7), and found managing windows to be very \textit{flexible} (4).
Notably, no specific navigation issues were identified while performing tasks in \vrBranchT{}.


\section{Findings and Discussions}
\label{sec:discussion}

Our primary objective is to validate the advantages of \VRT{} over the conventional desktop computing environment. 
While there's a growing interest in integrating VR/AR into data analytics, as highlighted in a recent state-of-the-art report~\cite{ens2021grand}, empirical studies directly contrasting VR with traditional environments remain limited. 
Accordingly, our research seeks to offer empirical insights into both the performance and user experience distinctions between \VRT{} and \DesktopT{} in the context of computational notebooks. 
We also aim to present quantitative data regarding the efficacy of our refined ``branch \& merge'' design. 
Our subsequent findings and discussions will center on these two focal areas.

\subsection{Is \VRT{} beneficial?}
Our study results reveal that \VRT{} markedly surpassed \DesktopT{} in the navigation task. However, this advantage did not extend to the comparison task, primarily due to \VRT{}'s inefficient text input mechanism.

\textbf{In terms of \textit{navigation},} participants operating within \VRT{} completed tasks faster than those using \DesktopT{}, both for deletion (one-stop navigation) and relocation (two-stop navigation). 
During one-stop navigation, participants scanned and traversed the notebook windows to locate a target. 
In the \VRT{} environment, this often involved head rotation, whereas in the \DesktopT{} environment, mouse scrolling was performed. 
Data suggests that physically rotating the head is more efficient than employing a mouse for navigating expansive information spaces that exceed the display size. 
This finding aligns with a study by Ball et al.~\cite{ball_move_2007}, which demonstrated the efficiency of physical navigation.

We also intentionally tested a two-stop navigation, where we still found \VRT{} to be faster compared to \DesktopT{}.
Participants, during this portion of the task, first located a target and subsequently repositioned it.
We anticipated that the consistent spatial environment in \VRT{} enhances participants' spatial recall, allowing them to identify the second target more efficiently than in \DesktopT{}. This would lead to a relatively smaller increase in completion time and ratio. 
Our reasoning aligns with prior research that examined the efficacy of spatial memory within \VRT{}~\cite{yang_virtual_2021,krokos2019virtual}.

\textit{In summary, for navigation,} we found that \VRT{} outperformed \DesktopT{} within computational notebooks, largely attributed to its accelerated browsing speed and enhanced spatial awareness and memory.

\textbf{In terms of \textit{comparison},} however, the pattern shifted. 
While \desktopLinearT{} and \vrLinearT{} exhibited comparable completion times, \vrBranchT{} lagged behind \desktopBranchT{}. 
As delineated in \autoref{sec:results-quan}, we foresaw—and subsequently confirmed—that challenges associated with text input considerably impacted the performance under \VRT{} conditions. 
After accounting for text input durations, \vrLinearT{} significantly outperformed \desktopLinearT{}, utilizing its navigational strengths. 
In the \LinearT{} conditions, the effectiveness of comparisons was closely tied to navigation efficiency since participants could view only one result at a time. 
However, the physical navigation capabilities of \VRT{} had a positive effect on comparison tasks in these conditions.
For the \BranchT{} conditions, the use of an intuitive \VRT{} gesture for creating branches was designed to make the process easier. 
Additionally, the larger display area offered by \VRT{} allowed for the simultaneous viewing of all results, which expedited the process of visual evaluation.
Yet, \vrBranchT{} only slightly edged out \desktopBranchT{}, not fully capitalizing on its navigational strengths.

Delving deeper into underlying factors, we concentrated on elements of the task not inherently tied to navigation. 
A detailed analysis of our interaction data unexpectedly revealed that participants took, on average, twice as long to generate a branch in \VRT{} compared to \DesktopT{}, see \autoref{fig:quant_findings_comparison}(e and f).
We hypothesize that this increased duration is attributable to the extended physical movements required by \VRT{}'s gestural interface, a notion supported by Fitts's law given the greater overall movement distance in the \VRT{} environment.
Additionally, we observed specific behavioral patterns among participants using \VRT{}. 
Typically, they would first grasp the window, retreat a step to initiate the branching process, and then advance to position the resultant windows. 
This step-back-and-forward motion appears to be a deliberate strategy to prevent the new branch windows from colliding with existing notebook windows, while also maintaining a consistent depth of window placement in space.
These additional interactions and subsequent fine-tuning of window positions incurred further time penalties in \VRT{}. 
Such findings are consistent with previous studies that have reported similar observations~\cite{bach_hologram_2018, wagner2018immersive,in2023table}.
Nevertheless, it's worth noting that despite the additional time required for the comparison task, participants expressed a clear preference for \VRT{}'s embodied interaction design, which received the highest ratings for both overall user experience and engagement.

\textit{In summary, for comparison,} \VRT{} maintains its navigational advantage, as navigation remains a crucial element in this task. 
However, the system's inefficiency in text input negatively impacted its overall performance. 
Additionally, although the embodied gesture interface in \VRT{} enhances user experience, it comes at the cost of increased task completion time.

\begin{figure}
    \centering
    \includegraphics[width=0.85\columnwidth]{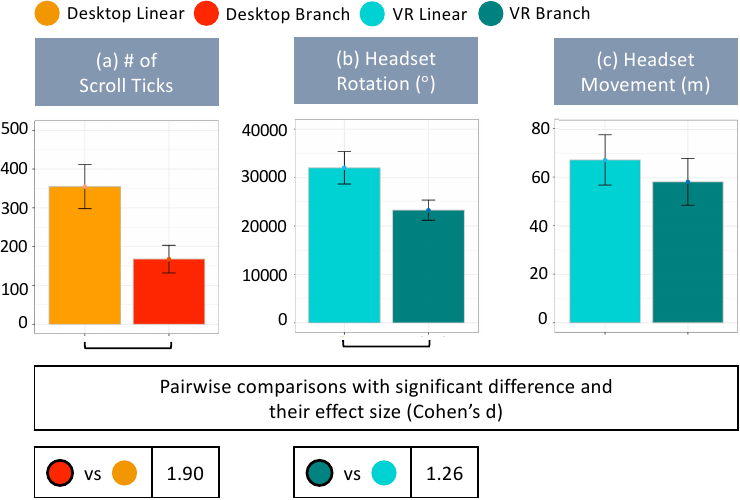}
    \caption{Navigation distance of different navigation methods across \DesktopT{} and \VRT{}: (a) the number of scroll ticks on the \DesktopT{}, (b) the degree of head rotation, and (c) the distance traversed in \VRT{}. Solid lines indicate statistical significance with $p < 0.05$. The tables below show the effect sizes for pairwise comparison. Circles with black borders indicate less navigation distance required.}
    \Description{Bar charts show the quantitative results of navigation counts they used in each desktop and VR environment.   The bar chart has an error bar showing 95\% confidence intervals. The bottom table shows the effect sizes of significant comparisons. Full Details are in Section 6.2.}
    \label{fig:quant_findings_comparison}
    \label{fig:nav-in-comparison}
\end{figure}

\subsection{Is ``branch \& merge'' beneficial?}
Yes, in our analysis, we found that the introduction of the \BranchT{} feature considerably shortened the completion time for comparison tasks in both \VRT{} and \DesktopT{}. 
Additionally, participants evidently perceived the \BranchT{} feature positively in terms of \textit{mental demand}, \textit{engagement}, and \textit{effectiveness}, as well as the \textit{overall user experience ranking}.
This subjective feedback aligns closely with findings from Weinman et al.~\cite{weinman2021fork}, wherein participants rated a similar desktop ``branch'' implementation. 
Our contributions extend this understanding by providing quantitative measures: for the comparison task in our study, the ``branch \& merge'' feature nearly reduced half of the completion time compared to its absence. 
Although creating a branch may initially take extra time, incorporating features like merging after branching can significantly reduce visual clutter and organize results spatially. 
This approach greatly decreases the amount of navigation effort needed.
Further post hoc analysis of the navigation methods in \DesktopT{} and \VRT{} validated our observations: both mouse scrolling and head rotation distances were notably shorter in \BranchT{} than in \LinearT{} as shown in \autoref{fig:nav-in-comparison}. 
To conclude, the ``branch \& merge'' feature enhances the comparison process, and our study presents no evident drawbacks related to its use.

\section{Generalizations, Limitations, and Future Work}
\label{sec:limitations}

\textbf{Generalizations.}
Regarding \textit{navigation}, we consider that the benefits we identified from physical navigation in \VRT{} may extend to a wide range of applications in immersive environments. 
This is attributable to the native support for head rotation and physical walking by the spatial tracking capabilities of VR/AR platforms. 
Our findings appear particularly relevant to VR/AR applications where users interact with multiple spatially-arranged windows, such as documents~\cite{davidson2022exploring}, images~\cite{luo2022should}, data tables~\cite{in2023table}, maps~\cite{satriadi2020maps}, and a mixed of applications~\cite{VisionPro}.
For more egocentric experiences, where a user is in a singular, immersive scene~\cite{yang2020embodied,yalong_yang_maps_2018,kraus_impact_2019}, our insights on navigation could still retain some relevance. 
However, the potential for increased occlusion in these views calls for further research.

Concerning \textit{comparison}, the ``branch \& merge'' method proved effective in both Desktop and VR environments, with the potential for beneficial integration into data flow frameworks{~\cite{yu2016visflow, yu2019flowsense}}, no-code platforms{~\cite{DataRobot, KNIME}}, and interactive visual programming{~\cite{du2023rapsai}}.
These systems typically employ a graph metaphor, wherein nodes denote data or functions, and links bridge the output of one node to the input of another. 
This metaphor, mirroring the interconnected windows in our computational notebooks, naturally supports the essential ``branch \& merge'' principles of code reuse maximization and simplifying logic.
As parameter spaces evolve in complexity, the ``branch \& merge'' functionality holds significant promise in facilitating hypothesis testing.


\textbf{Text interaction in VR.}
Our study underscored challenges associated with text interactions in VR, encompassing issues like text selection, defining the entry point, and the actual typing process.
On an optimistic note, the significance of enhancing text interaction for VR productivity tasks has gained consensus, and as such, it's an evolving research domain~\cite{knierim2020opportunities,dube2019text}. 
Several innovative solutions tailored for stationary environments have been presented, such as tracking physical keyboards~\cite{MetaKeyboard} or emulating keyboards on flat surfaces. 
However, these challenges amplify when one introduces movement within the VR space.
Advancements in sensory technologies and hardware, including haptic gloves and enhanced finger tracking, are likely to refine the VR text interaction experience in dynamic settings in the foreseeable future.
An interim solution might involve minimizing mandatory text interactions. Employing input widgets, such as dropdown menus and sliders---features commonly found in data flow systems~\cite{ApacheNiFi, ApacheCamel} or no-code data science tools~\cite{ORANGE, RapidMiner}---could serve this purpose.

\textbf{Embodied gesture in VR.}
Our study revealed that while embodied gestures in VR enhanced the user experience, they also necessitated a longer execution time. 
Several factors might account for this extended duration: the greater movement distance, the need for precise placement adjustments, and efforts to prevent interference with other visual elements. 
The latter two challenges present opportunities for improvement. 
To address issues related to hand tremors or shaky mid-air gestures, future developments could incorporate a proximity snapping technique that automatically aligns the window to a predefined position as it approaches a designated area~\cite{seth2011virtual}. 
Additionally, participants in our study appeared to avoid element collisions unconsciously; future gesture design should consider this behavior. For instance, a branching gesture could be executed orthogonally to the document layout in the depth direction, thereby minimizing the risk of collision with adjacent windows. Subsequent research should validate these observations and contribute to the development of systematic guidelines for gesture design in VR.


\textbf{Scalability.}
In a recent analysis of publicly accessible Jupyter notebooks (N=470), the study found that the average notebook comprised 125 lines of code and 20 cells~\cite{ramasamy2023workflow}.
The notebooks evaluated in our study were of comparable lengths. 
When considering the accommodation of longer notebooks, several potential strategies emerge. 
One approach is to extend the curvature of the layout, positioning notebooks at a greater distance from the user. 
However, this requires increased user movement and could compromise content readability. 
Alternatively, vertical space could be utilized to arrange windows in a grid format; however, this introduces the challenge of accessing elevated windows and may require additional interaction designs.
In summary, future research should explore these trade-offs and consider other potential solutions for effectively accommodating longer computational notebooks and more complex branching scenarios.

\textbf{Addressing Additional Computational Notebook Challenges in VR.}
Our study primarily aims to leverage VR for enhancing navigation and comparison in computational notebooks, as these are fundamental interactions in data analysis where VR can potentially offer significant improvements. 
We acknowledge that our design may not represent the optimal adaptation of the computational notebook framework in VR. 
Our current adaptation focuses on examining the impacts of specific factors we aimed to explore, but other innovative approaches could exist.
Moreover, computational notebooks face various other challenges, like the ones identified in a previous comprehensive study: setup, exploration and analysis, managing code, reliability, archival, security, sharing and collaboration, reproducing and reusing, and notebooks as products~\cite{chattopadhyay2020s}.
We believe addressing those challenges will be a long-term effort from multiple communities. For example, Wang et al.~\cite{wang2020callisto} highlighted the possibilities for real-time collaboration among multiple users. 
Moving forward, we want to explore how to better exploit the unique display and interaction capabilities of VR to improve those experiences.

\section{Conclusion} 
\label{sec:conclusion}

We adapted the computational notebook interface from desktop to VR and tested its effectiveness through a controlled study. 
Our results revealed that notebooks in \VRT{} outperformed notebooks on \DesktopT{} in navigation efficiency, and the inclusion of a  ``branch \& merge''  feature notably enhanced the non-linear comparison process. 
Participants reported that the integration of VR with the "branch\&merge" functionality was the most engaging and provided the best overall user experience among all test conditions.
However, we observed that text interaction in VR remains a challenge. This issue could be alleviated with future advancements in hardware and tracking technologies, and as users become more familiar with VR environments over time. 
Our study underscores the immense potential of computational notebooks in VR, particularly in enhancing navigation and comparison performance and experience for analysts.
It's important to note that our VR adaptation was specifically tailored to investigate navigation and comparison, and there may be other innovative approaches for adapting or completely redesigning computational notebooks in VR. 
Broadly, our results provide preliminary evidence supporting the wider use of large display spaces, augmented spatial awareness, embodied interaction, and physical navigation in VR for immersive analytics applications.

\begin{acks}
This work was supported in part by NSF I/UCRC CNS-1822080 via the NSF Center for Space, High-performance, and Resilient Computing (SHREC).
\end{acks}

\bibliographystyle{ACM-Reference-Format}
\bibliography{source/main}

\end{document}